\documentstyle[12pt,epsfig]{article}

\newcommand{\bq}{\begin{equation}}
\newcommand{\fq}{\end{equation}}
\newcommand{\bqr}{\begin{eqnarray}}
\newcommand{\fqr}{\end{eqnarray}}

\newcommand{\xpv}[1]{\langle #1  \rangle}

\newcommand{\rf}[1]{(\ref{#1})}

\def\jhep#1#2#3{J. High Energy Phys. {\bf #1} (#2) #3}

\def\alp{\alpha}   \def\bet{\beta}    
\def\del{\delta}    
    \def\th{\theta}

 \def\vphi{\varphi}

  \def\cL{{\cal L}}
  \def\cO{{\cal O}}

\def\pr{^{\prime}}

\def\rar{\rightarrow}

\newcommand{\tr}{\mbox{Tr}}

\def\ove#1{\frac{1}{#1}}

\begin{document}

\thispagestyle{empty}

\marginparwidth = .5in

 \marginparsep = 1.2in

\begin{flushright}
\begin{tabular}{l}
ITP-SB-98-65    \\  hep-th/9811155
\end{tabular}
\end{flushright}

\vspace{18mm}
\begin{center}

{\bf A note on four-point functions of conformal operators in $N=4$ Super-Yang Mills.}

\vspace{18mm}

{Francisco Gonzalez-Rey}\footnote{E-mail
address:glezrey@insti.physics.sunysb.edu},
 {Inyong Park}\footnote{E-mail
address:ipark@insti.physics.sunysb.edu}
, Koenraad Schalm\footnote{E-mail
address:konrad@insti.physics.sunysb.edu} \\[10mm] {\em Institute for
Theoretical Physics
\\ State University of New York       \\ Stony Brook, NY 11794-3840, USA }

\vspace{20mm}

{\bf Abstract}
\end{center}
We find that the first-order correction to the free-field result for the
four-point function of the conformal operator $\tr(\phi^i\phi^j)$ is nonvanishing and survives in the limit $N_c \rar \infty$.

\vfill
PACS: 11.30.Pb 11.40
\newpage
\setcounter{footnote}{0}
\setcounter{page}{1}

\noindent
{\bf I. Introduction and conclusion}
\vspace{.2in}

\noindent
Since the conjecture by Maldacena that gauged supergravity on an
anti-de Sitter background is dual to the large $N_c$, large
$g^2_{YM}N_c$ limit of conformal field theories, many
efforts have been made to test the conjecture as well as to extract predictions
about the behaviour of the conformal field theories in this limit. In the last
class falls the recent conjecture by Intriligator $\cite{I}$ that the
S-duality of type IIB supergravity on $\mbox{AdS}_5 \times S_5$ constrains the
correlation functions of conformal operators $\cO_i$ in short
representations of the conformal group of the dual CFT: 
$d=4$
$N=4$ super-Yang-Mills. In the limit of small $g_{YM}$ and large $g^2_{YM}N_c$ they must obey the selection rule
\bq
\xpv{\prod_{i=n}^n \cO^{(q_i)}_i(x_i)} =0 ~~~\mbox{unless}~~~ \sum_i q_i =0
\label{int}
\fq
The $q_i$ are the charges of a $U(1)_Y$ bonus symmetry of the correlation functions, present in the double limit $g^2_{YM}N_c,~N \rar \infty$. This $U(1)_Y$ is the maximal compact subgroup of the $SL(2,Z)$ S-duality group.

An attempt to prove the validity of this selection rule  using harmonic $N=4$ superspace in the last section of the paper \cite{I} found the suspicious result that this symmetry should hold for all values of $g^2_{YM}$ and $N_c$. This would in turn imply that all correlation functions are given by their free field values. In this brief note we merely want to point out that it is very easy to construct a counter-example. This is perhaps a little superfluous as it was already pointed out in the literature  that $n \geq 4$ correlation functions receive $1/g^2_{YM}N$ corrections from the AdS point of view \cite{banks}. This should be enough evidence to preclude that the answer is free field theory. That $N=4$ SYM  is interacting is also evident from the instanton calculations in \cite{inst}. These corrections are subleading in $N_c$, however, and conceivably the theory could still simplify in the limit $N_c \rar \infty$. The calculation presented below, that the four-point function of the $\tr \phi^a \phi^b$ conformal operator receives planar diagram corrections at order $g^2_{YM}N_c$  corroborates the AdS result and strengthens the instanton calculations from the weak coupling point of view: $N=4$ SYM at large $N_c$ is interacting.
\vspace{.3in}

\noindent
{\bf II. Computation of the correlator $\xpv{(\tr\phi\phi)^4}$}
\vspace{.2in}

\noindent
We will perform our calculation in $N=1$ superspace. In $N=0$ components the
conformal operator $\tr \phi^a\phi^b$ transforms in the ${\bf 20}\pr$
representation of $R$-symmetry group $SU(4)_R \simeq SO(6)$. The $\phi^a$ are
the six real scalar fields of $N=4$ SYM transforming in the adjoint of the
gauge group $SU(N_c)$; $a$ is an $SO(6)$ vector index. In $N=1$ superfields
these operators belong to the three supermultiplets $\tr \Phi_i\Phi_j$, $\tr
\Phi_i e^V \bar{\Phi}^{{j}}$, $\tr \bar{\Phi}^{{i}}
\bar{\Phi}^{{j}} $ transforming in the ${\bf 6}$, ${\bf 8}$ and
${\bf \bar{6}}$ of the manifest $SU(3)$ flavor symmetry of $N=4$ SYM in $N=1$
language. The $e^V$ in $\tr
\Phi_i e^V \bar{\Phi}^{{j}}$ is necessary for gauge invariance of the operator.\footnote{One may of course covariantize the antichiral field $\bar{\Phi}$ to a ``vector'' representation $\bar{\Phi} \rar e^V\bar{\Phi}$, in which case the operators look more symmetric.}

The $N=1$ Lagrangian is
\bqr
\cL &=& -\int d^2\th \frac{1}{2} \tr W^{\alp}W_{\alp} - \int d^2\th\,\tr \Phi_1[\Phi_2,\Phi_3]
+ \int d^2\bar{\th} \,\tr \bar{\Phi}^1[\bar{\Phi^2},\bar{\Phi^3}] \\ \nonumber 
&&\hspace{3in}
+ \int d^2\th d^2\bar{\th} \,\tr \Phi^i e^{V}\bar{\Phi}_i ~,
\label{act}
\fqr
and we include a source term for the conformal operators
\bq
\cL_{source} = \int d^2\th J^{ij}\tr \Phi_i\Phi_j + \int d^2\bar{\th}
\bar{J}_{ij}\tr\bar{\Phi}^i\bar{\Phi}^j ~.
\fq
The field strength equals $W_{\alp} = \bar{D}^2e^{-V}D_{\alp}e^V$ in terms of
the prepotential $V$. All fields are in the adjoint representation of the gauge
group $SU(N_c)$ and we have chosen antihermitian generators, i.e. $\tr T^AT^B =-\del^{AB}C_2(\mbox{Ad})=-\del^{AB}N_c$. The full Lagrangian, including
the source term, $\cL+\cL_{source}$, is multiplied by the inverse of the
coupling constant squared $1/g^2_{YM}N_c$. This corresponds to the normalization
of the conformal operator $\cO_{ij}= \frac{1}{g^2_{YM}N_c}\tr\Phi_i\Phi_j$ where
the free-field result is $g^2_{YM}N_c$ independent.

It will suffice for our purposes to compute the
single correlator $\xpv{\cO_{11}\cO_{22}\bar{\cO}^{11}\bar{\cO}^{22}}$ and show that its
first-order correction is nonzero. The connected free-field result for this correlator, given by figure 1,
vanishes as all the propagators are diagonal in the flavor group $SU(3)$.
\begin{figure}[ht]
\begin{center}
{\epsfxsize=2in \epsfbox[110 150 450 440]{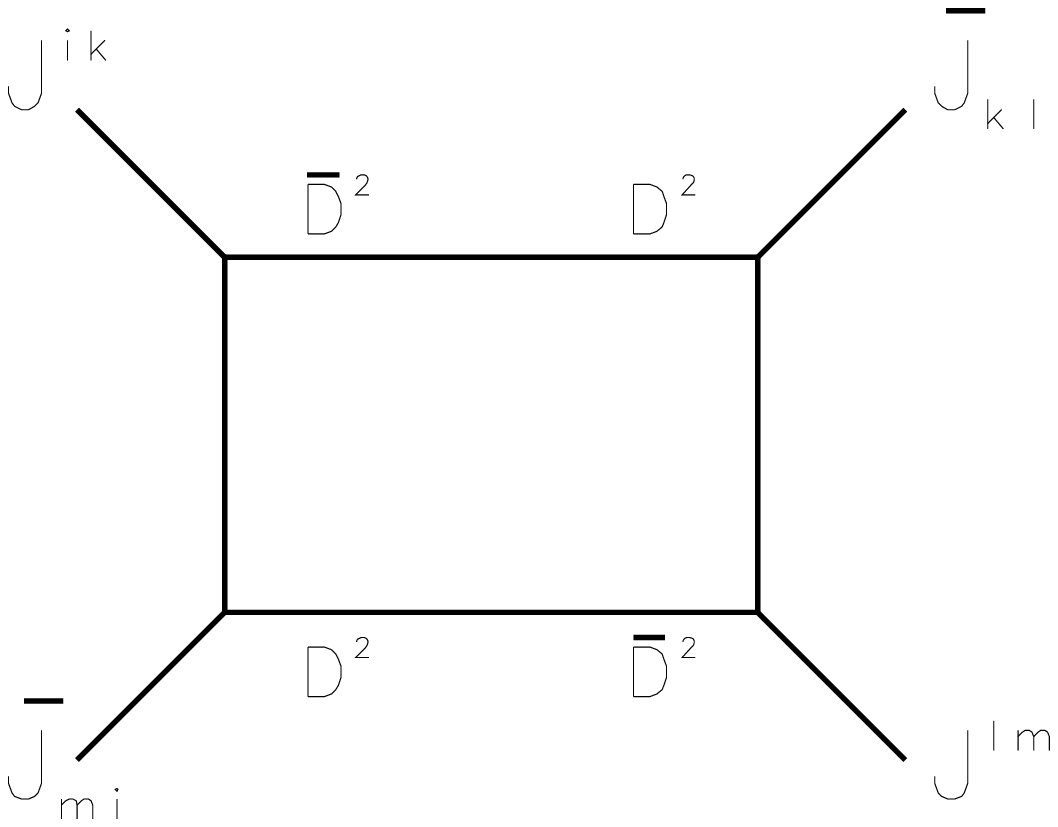}}
\end{center}
\caption{Lowest order contribution to $\xpv{(\tr\phi\phi)^4}$}
\end{figure}

The first-order corrections are given
by Feynman diagrams of the form given in figure 2. We may ignore self-energy corrections to the propagator as in $N=4$ there is no wave-function renormalization at one-loop. In principle one should also include all the connected but not 1PI diagrams. However, these graphs vanish by virtue of the tracelessnes of the
$SU(N_c)$ structure constants. Moreover, the couplings to the vector multiplet are diagonal in
flavors and therefore the graphs in figures 2b, 2c and 2d again vanish trivially.
\begin{figure}[ht]
\begin{center}
\begin{tabular}{cc}
\raisebox{-.28in}{\epsfxsize=2in \epsfbox[110 100 450 440]{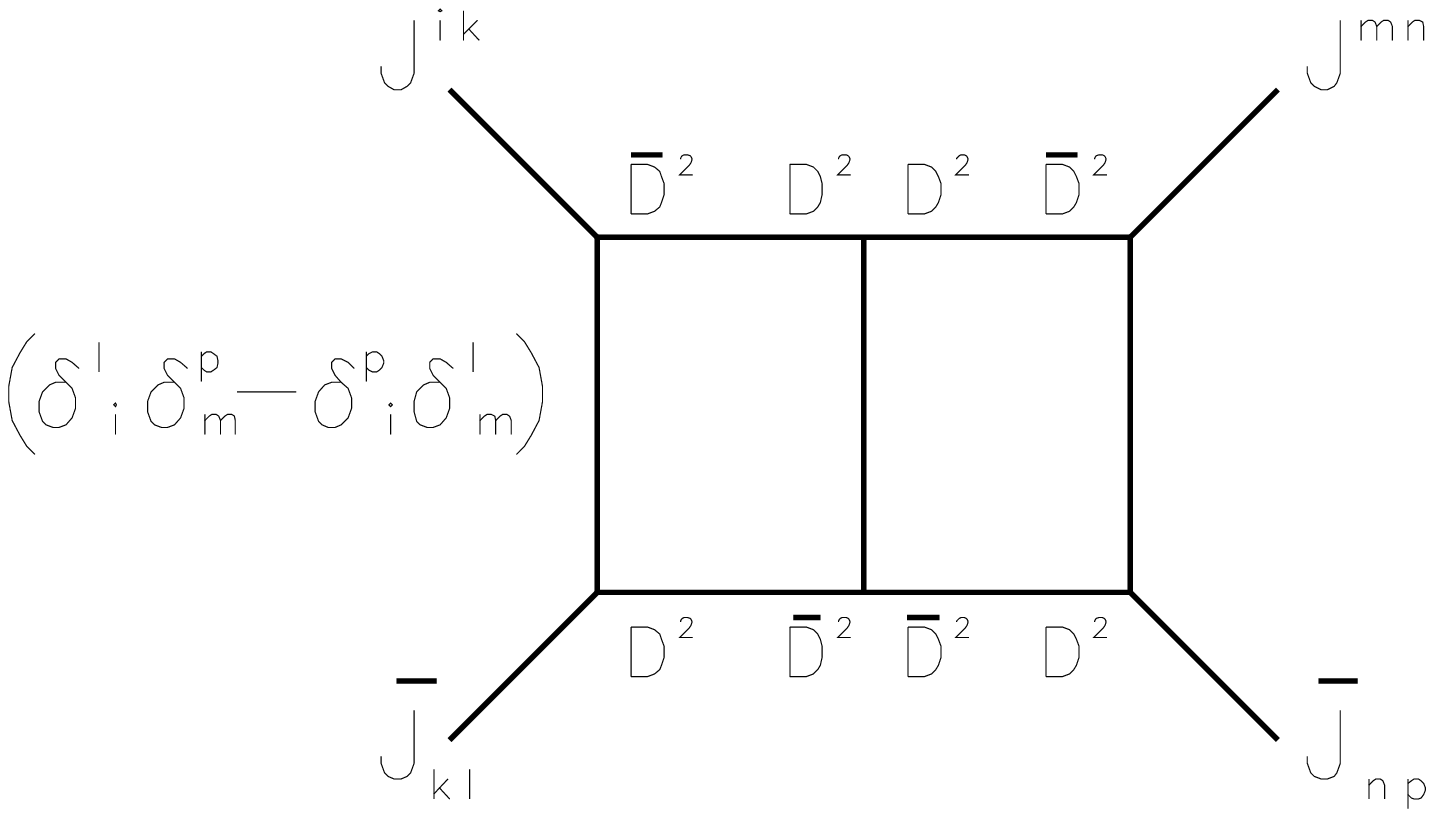}} & {\epsfxsize=2in \epsfbox[110 150 450 440]{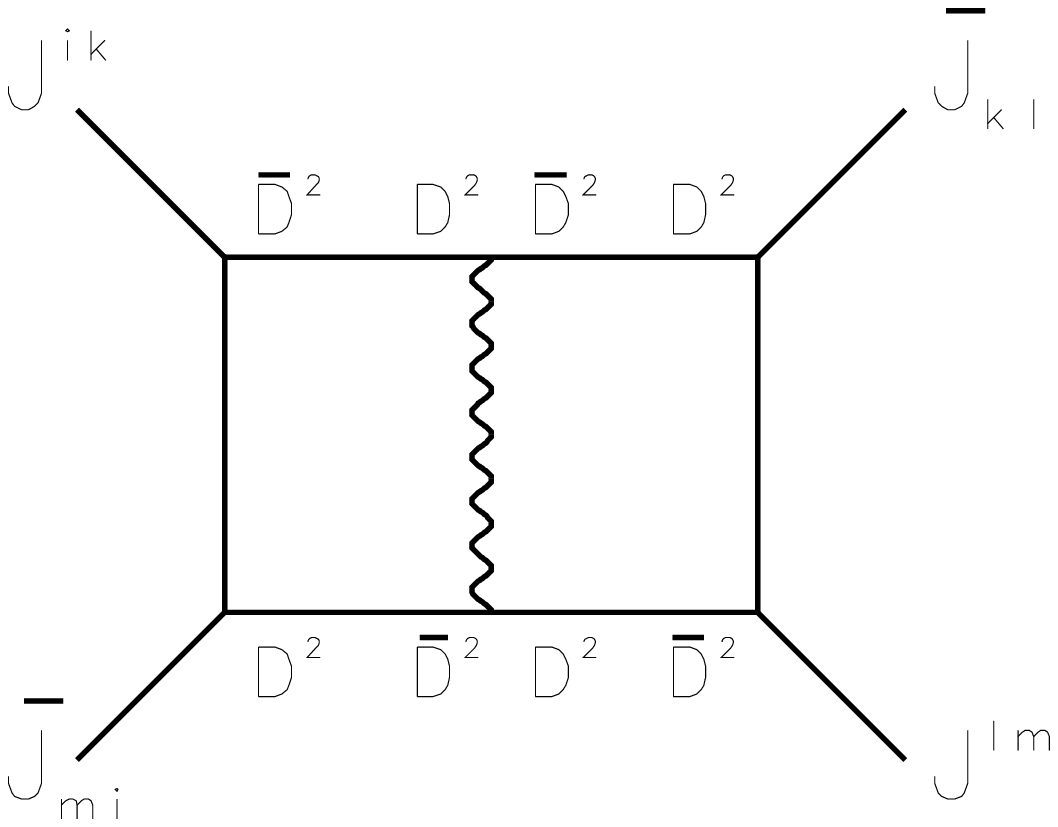}}  \\
2a. & 2b.   \\
{\epsfxsize=2in \epsfbox[110 150 450 440]{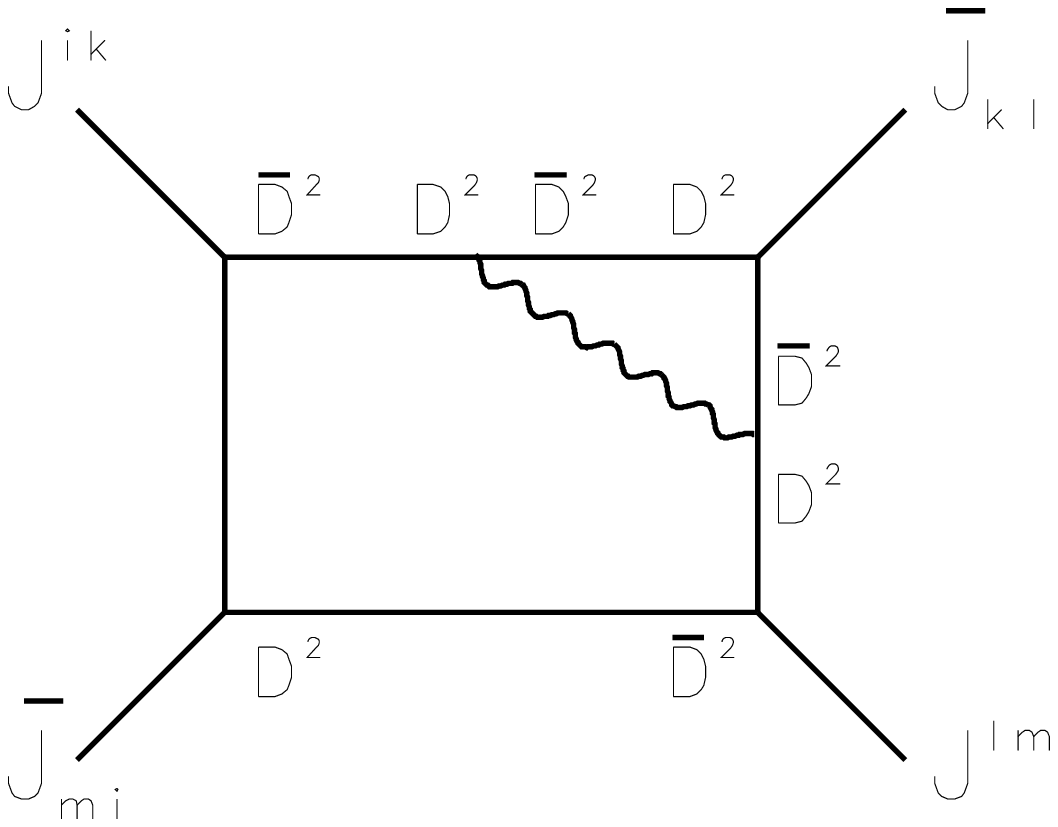}}& {\epsfxsize=2in \epsfbox[110 150 450 440]{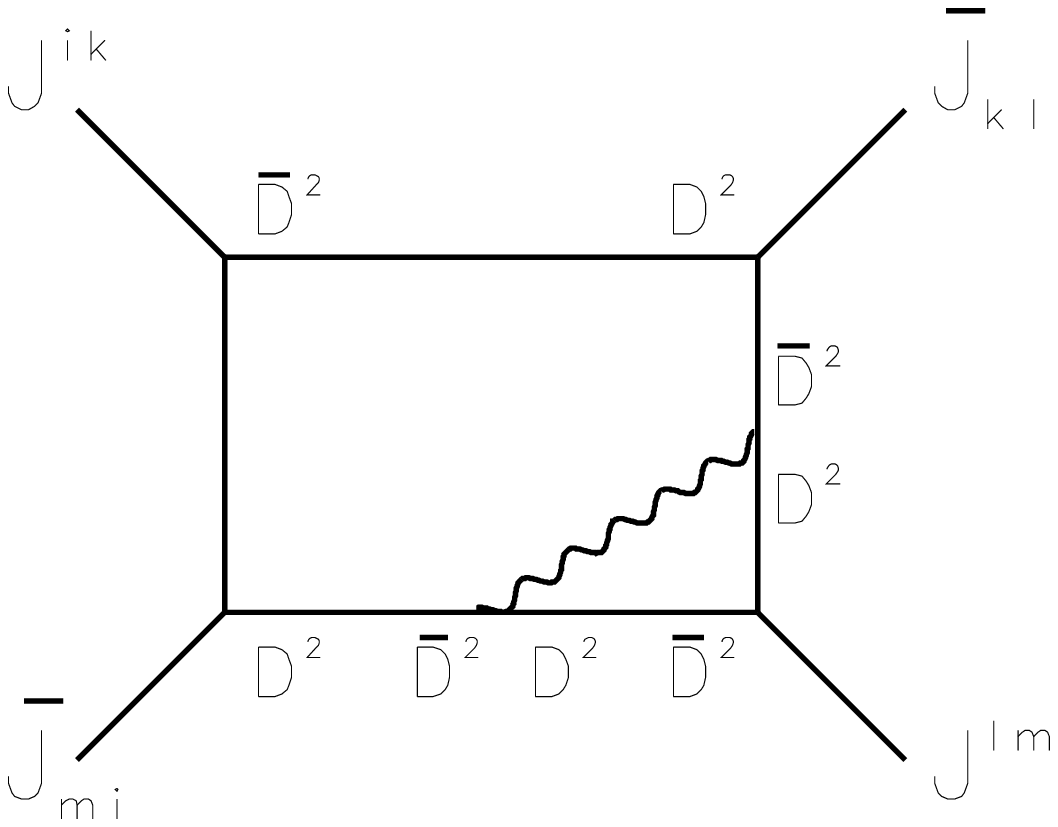}}  \\
2c. & 2d. 
\end{tabular} 
\end{center} 
\caption{Feynman diagrams contributing to $\xpv{(\tr\phi\phi)^2}$.}
\end{figure}
Thus the sole graph in figure 3a remains to be
evaluated. 
\begin{figure}[ht]
\begin{center}
\begin{tabular}{cc}
{\epsfxsize=2in \epsfbox[110 150 450 440]{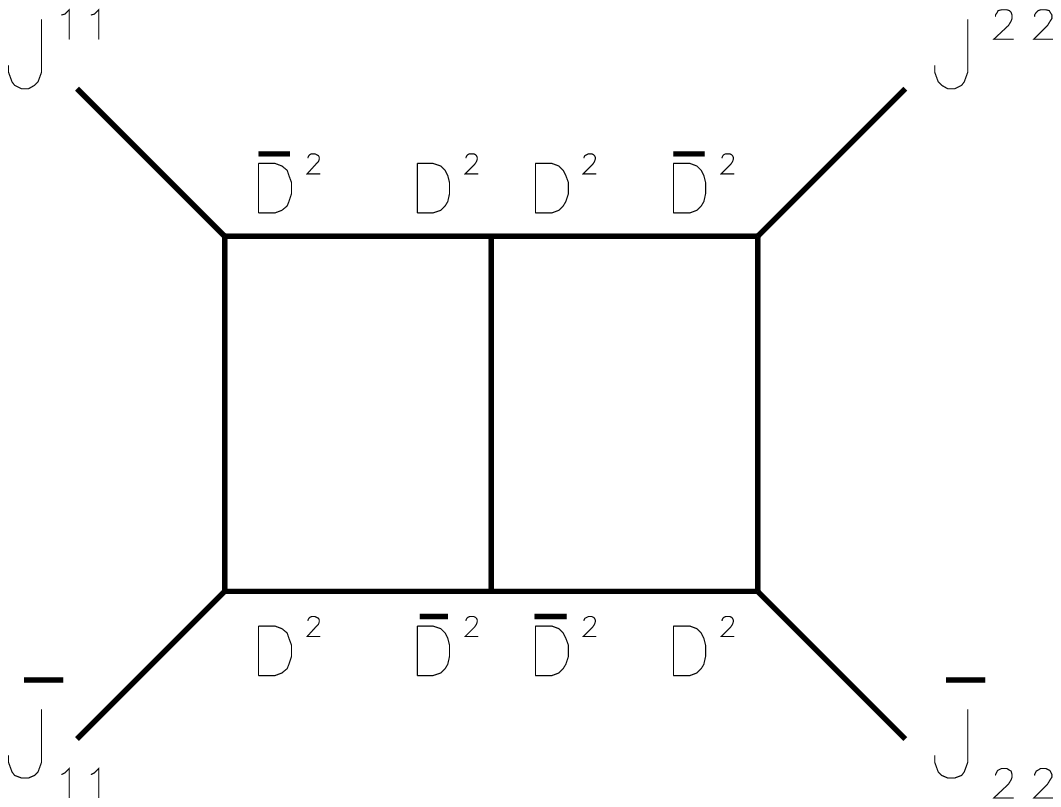}} 
&
{\epsfxsize=2in \epsfbox[110 150 450 440]{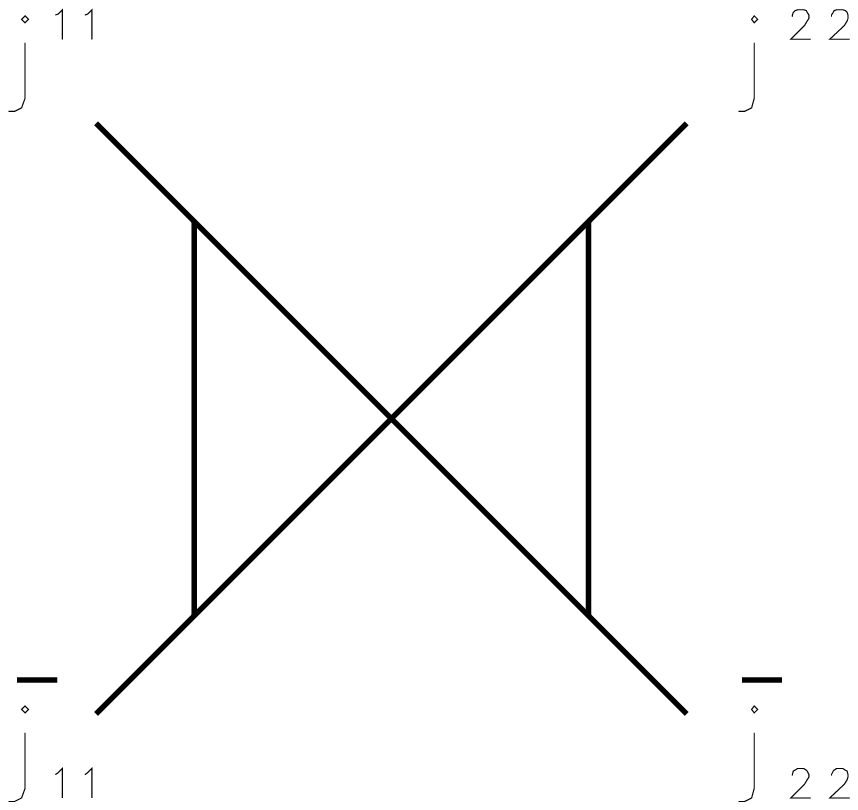}} \\
 3a. & 3b.
\end{tabular}
\end{center}
\caption{Sole contribution to the correlator
$\xpv{\cO_{11}\cO_{22}\bar{\cO}^{11}\bar{\cO}^{22}}$ in superfields and components}
\end{figure}

The effective action for the sources $J^{ij}$ and $\bar{J}_{ij}$ which is responsible for this diagram is given by
\bqr
\nonumber
S_{eff} &=& -\frac{16g_{YM}^2N_c}{C_2(\mbox{Ad})}f_{abc}f^{abc}\int \prod_{i=1}^6 d^8z_i ~\frac{\del^8_{13}}{\Box}\frac{D^2\bar{D}^2\del^8_{16}}{\Box}\frac{D^2\bar{D}^2\del^8_{35}}{\Box}\frac{\del^8_{56}}{\Box}\frac{D^2\bar{D}^2\del^8_{26}}{\Box}\frac{\del^8_{24}}{\Box}\frac{D^2\bar{D}^2\del^8_{45}}{\Box}
\\ 
&&\hspace{2.6in} \times J^{11}(z_1)J^{22}(z_2)\bar{J}_{11}(z_3)\bar{J}_{22}(z_4) ~.
\fqr
We perform the $D$-algebra such that only terms that have the structure $J \bar{J}D^2J\bar{D}^2\bar{J}$ contribute to the scalar components of $\cO$ and keep only these terms. We find
\bqr
S_{eff} &=& 16g_{YM}^2N_c(N_c^2-1)\int \prod_{i=1}^6 d^8z_i ~\frac{\del^8_{13}}{\Box}\frac{\del^8_{16}}{\Box}\frac{\del^8_{35}}{\Box}\frac{\Box\del^8_{56}}{\Box}\frac{\del^8_{26}}{\Box}\frac{\del^8_{24}}{\Box}\frac{\del^8_{45}}{\Box}
\label{str}
\\ \nonumber &&\hspace{2in} \times
J^{11}(z_1)D^2J^{22}(z_2)\bar{J}_{11}(z_3)\bar{D}^2\bar{J}_{22}(z_4) \\ \nonumber
&& \hspace{2.2in} + \ldots
\fqr
The Grassmann integrations may be performed and as usual the final expression is local in Grassmann space. For $N_c \gg 1$, $g^2_{YM}N_c$ fixed, the result translates into the nonvanishing first-order correction to the correlation function

\bqr \nonumber
\xpv{\cO_{11}(z_1)\cO_{22}(z_2)\bar{\cO}^{11}(z_3)\bar{\cO}^{22}(z_4)}|_{\th=0} && \\
&& \hspace{-2.5in}= \frac{16g_{YM}^2N_c^3}{(2\pi)^8}\ove{x_{13}^2x_{24}^2}\int d^4x_a \ove{x_{1a}^2}\ove{x_{2a}^2}\ove{x_{3a}^2}\ove{x_{4a}^2}
\label{res}
\\ \nonumber
&&\hspace{-2.5in} =\frac{g_{YM}^2N_c^3}{2^4 \pi^6}\frac{1}{x_{12}^2x_{13}^2x_{24}^2 x_{34}^2} \int_0^{\infty} \frac{d\bet}{\bet} {}_2F_1(1,1;2;1-\frac{(x_{13}^2+\bet x_{23}^2)(x_{14}^2+\bet x_{24}^2)}{\bet x_{12}^2x_{34}^2}) 
\fqr
where ${}_2F_1(1,1;2;1-y)= -\ln(y)/(1-y)$ is a special case of the hypergeometric function. Eq.\rf{res} is the expected result for a four-point correlator of conformal operators \cite{fran}.

From eq.\rf{str} one sees that the component expression for the correlator of $\cO_{ij}|_{\th=0} = \tr\vphi_i\vphi_j$ ($\vphi_i$ is a complex combination of the six $\phi^a$) receives only a contribution from the exchange of auxiliary fields. Integrating these out in the component form of the action~\rf{act} yields the effective interaction
\bq
S^{(N=0)}_{int} = \ldots + \tr[\phi_i,\bar{\phi}^i][\phi_j,\bar{\phi}^j]-2\tr[\phi_i,\phi_j][\bar{\phi}^i,\bar{\phi}^j]
\fq
This interaction is responsible for the diagram in figure 3b. 
One sees immediately that its contribution is exactly that of eq.~\rf{res}.

As a last comment we should emphasize that because the free-field result is vanishing, the conclusion that $N=4$ SYM at large $N_c$ is interacting is robust. The corrections cannot be compensated by redefining or re-normalizing the operators.
\vspace{.3in}

\noindent
{\bf Acknowledgements}
\vspace{0.1in}

\noindent
We are grateful to  M. Ro\v{c}ek for enlightening discussions.

\end{document}